\documentclass[11pt]{article}
\usepackage{amssymb,latexsym,amsmath}
\usepackage{amsfonts, graphics}
\newtheorem{theorem}{Theorem}
\newtheorem{lemma}{Lemma}

\def\open#1{\setbox0=\hbox{$#1$}
\baselineskip = 0pt
\vbox{\hbox{\hspace*{0.4 \wd0}\tiny $\circ$}\hbox{$#1$}}
\baselineskip = 11pt\!}

\begin{document}
\title{Sharp bounds on $2m/r$ of general spherically symmetric static objects}
\author{H{\aa}kan Andr\'{e}asson\\
        Mathematical Sciences\\
        Chalmers and G\"{o}teborg University\\
        S-41296 G\"oteborg, Sweden\\
        email\textup{: \texttt{hand@math.chalmers.se}}}

\maketitle

\begin{abstract}
In 1959 Buchdahl \cite{Bu} obtained the inequality $2M/R\leq 8/9$
under the assumptions that the energy density is non-increasing
outwards and that the pressure is isotropic. Here $M$ is the ADM
mass and $R$ the area radius of the boundary of the static body.
The assumptions used to derive the Buchdahl inequality are very
restrictive and e.g. neither of them hold in a simple soap bubble.
In this work we remove both of these assumptions and consider
\textit{any} static solution of the spherically symmetric Einstein
equations for which the energy density $\rho\geq 0,$ and the
radial- and tangential pressures $p\geq 0$ and $p_T,$ satisfy
$p+2p_T\leq\Omega\rho, \;\Omega>0,$ and we show that
$$\sup_{r>0}\frac{2m(r)}{r}\leq
\frac{(1+2\Omega)^2-1}{(1+2\Omega)^2},$$ where $m$ is the
quasi-local mass, so that in particular $M=m(R).$ 
We also show that the inequality is sharp. Note
that when $\Omega=1$ the original bound by Buchdahl is recovered.
The assumptions on the matter model are very general and in
particular any model with $p\geq 0$ which satisfies the dominant
energy condition satisfies the hypotheses with $\Omega=3.$
\end{abstract}

\section{Introduction}
\setcounter{equation}{0}
The metric of a static spherically symmetric spacetime takes the following form in Schwarzschild coordinates
\begin{displaymath}
ds^{2}=-e^{2\mu(r)}dt^{2}+e^{2\lambda(r)}dr^{2}
+r^{2}(d\theta^{2}+\sin^{2}{\theta}d\varphi^{2}),
\end{displaymath}
where $r\geq 0,\,\theta\in [0,\pi],\,\varphi\in [0,2\pi].$
Asymptotic flatness is expressed by the boundary conditions
\begin{displaymath}
\lim_{r\rightarrow\infty}\lambda(r)=\lim_{r\rightarrow\infty}\mu(r)
=0,
\end{displaymath}
and a regular centre requires $\lambda(0)=0.$
The Einstein equations read
\begin{eqnarray}
&\displaystyle e^{-2\lambda}(2r\lambda_{r}-1)+1=8\pi r^2\rho,&\label{ee1}\\
&\displaystyle e^{-2\lambda}(2r\mu_{r}+1)-1=8\pi r^2
p,&\label{ee2}\\
&\displaystyle
\mu_{rr}+(\mu_{r}-\lambda_{r})(\mu_{r}+\frac{1}{r})= 8\pi p_T
e^{2\lambda}.&\label{ee4}
\end{eqnarray}
Here $\rho$ is the energy density, $p$ the radial pressure and
$p_T$ is the tangential pressure. If the pressure is isotropic,
i.e., $p=p_T,$ a solution will satisfy the well-known
Tolman-Oppenheimer-Volkov equation for equilibrium
\begin{equation}
p_{r}= -\mu_{r}(p + \rho).\label{TOV1}
\end{equation}
In the case of non-isotropic pressure this equation generalizes to
\begin{equation}
p_{r}= -\mu_{r}(p + \rho) - \frac{2}{r}(p-p_T).\label{TOV2}
\end{equation}
Note that the radial pressure $p$ is monotone in the isotropic case 
if $p+\rho\geq 0$ since $\mu_r\geq 0,$ cf. (\ref{mur}). 
The quasi-local mass $m=m(r)$ is given by
\begin{equation}\label{m}
m(r)=\int_0^{r}4\pi\eta^2 \rho(\eta)d\eta,
\end{equation}
and the ADM mass of a steady state for which the energy density has support in $[0,R]$ 
is thus given by $M=m(R).$

Schwarzschild asked already in 1916 the question: How large can
$2M/R$ possibly be? He gave the answer $2M/R\leq 8/9$ \cite{S} in
the special case of the Schwarzschild interior solution which has
constant energy density and isotropic pressure. In 1959 Buchdahl
\cite{Bu} extended his result to isotropic solutions for which the
energy density is non-increasing outwards and he showed that also
in this case
$$2M/R\leq 8/9.$$

This is called the Buchdahl inequality and is included in most text books
on general relativity in connection with the discussion of the interior
solution by Schwarzschild, cf. e.g. \cite{Wa} and \cite{We}.
The quantity $2m/r$ is
fundamental for determining the spacetime geometry of a static
spherically symmetric spacetime, cf. equations (\ref{eminustwolambda}) and
(\ref{muandlambda}). A bound on $2M/R$ has also an
immediate observational consequence since it limits the possible
red shifts of spherically symmetric static objects. 

The assumptions made by Buchdahl are extremely restrictive as
pointed out by Guven and \'{O} Murchadha \cite{GM}, e.g. neither of
the assumptions hold in a simple soap bubble and they do not
approximate any known topologically stable field configuration.
Moreover, astrophysical models of stars are not unusually
anisotropic. Lemaitre proposed a model of an anisotropic star
already in 1933 \cite{Le}, and Binney and Tremaine \cite{BT} explicitly allow
for an anisotropy coefficient (cf. also \cite{HS} and the references therein). 

One motivation for this study has its roots in the numerical
investigation of the spherically symmetric Einstein-Vlasov (ssEV)
system \cite{AR2} which admits a very rich class of static
solutions. The \textit{overwhelming} number of these have neither
an isotropic pressure nor a non-increasing energy density, but nevertheless $2M/R$ 
is always found to be less than $8/9,$ cf. \cite{AR2}. There are 
sometimes arguments which claim that the monotonicity of $\rho$ is necessary 
for the stability of a steady state, cf. e.g. \cite{We}, but at least 
for Vlasov matter this is not the case by the results presented in \cite{AR1}.

In this work the problem of finding a sharp bound on $2m/r$ is solved in full
generality in the class of matter models which satisfy
\begin{equation}\label{Omega}
p+2p_T\leq\Omega\rho,\mbox{ where }\Omega,p \mbox{ and }\rho\mbox{ are non-negative.}
\end{equation}
We will show that
$$\sup_{r>0}\frac{2m(r)}{r}\leq\frac{(1+2\Omega)^2-1}{(1+2\Omega)^2},$$
for any static solution of the spherically symmetric Einstein
equations which satisfies (\ref{Omega}).
The class of matter models defined by (\ref{Omega}) is very
general. Indeed, a realistic matter model should satisfy the
dominant energy condition (DEC) which implies that $\rho\geq 0$
and that the inequality (\ref{Omega}) holds with $\Omega=3.$ The
remaining condition that $p$ is non-negative is a standard
assumption for most matter models in astrophysics. Moreover,
Vlasov matter satisfies the conditions in (\ref{Omega}) with
$\Omega=1.$ An interesting feature of Vlasov matter, in comparison
with a fluid model, is that no equation of state which relates the
pressure and the energy density has to be specified. For Vlasov
matter, $\rho, p$ and $p_T$ are all determined by a single density
function on phase space, cf. \cite{An3} and \cite{Rl} for more
information on Vlasov matter and the EV system.

The bound that we obtain for $2m/r$ is sharp in the sense
that an infinitely thin shell of matter, with $2m/r$
equal to the critical value, will satisfy a form of the generalized TOV 
equation which allows $\rho$ and $p_T$ to be measures ($p=0$
here). This is described in detail in the next section. It should
here be pointed out that for the ssEV system the results in
\cite{An2} show that there exist \textit{regular} static solutions
with the property that $2M/R$ takes values arbitrary close to
$8/9$ ($\Omega=1$ for Vlasov matter). These solutions do approach
the infinitely thin shell mentioned above as $2M/R\to 8/9.$ In
section 4 we give an analogy with a classical problem in
electrostatics (or equivalently in Newtonian gravity) which
shares the property that the maximizer is a measure at the boundary.
In the work by Buchdahl, the solution that maximizes $2M/R$ is
the Schwarzschild interior solution with constant energy density. 
This solution has the property that the pressure becomes
unbounded as $2M/R\to 8/9,$ and therefore the solution does not satisfy
the DEC and is not a realistic steady state. 

Before finishing this section with a review of previous results,
let us point out that the original motivation for investigating
the Buchdahl inequality in full generality comes from its possible
role in understanding the formation of trapped surfaces.
Christodoulou has obtained conditions which guarantee the
formation of trapped surfaces in the case of a scalar field
\cite{Ch1}, and this result is crucial for his proof of the weak-
and strong cosmic censorship conjectures \cite{Ch2}. For more
information on this see the introduction in \cite{An2}.

General investigations of the Buchdahl inequality have previously
been undertaken by Baumgarte and Rendall \cite{BR} and Mars,
Merc\`{e} Mart\'{i}n-Prats and Senovilla \cite{MMMPS}. These
studies concern very general matter models and they obtain the
bound $2m/r<1.$ This bound gives little information on the spacetime 
geometry since $\lambda\to\infty$ as $2m/r\to 1,$ and in
particular it gives no bound on the red shift of a static body. In
\cite{An1} shells supported in $[R_0,R_1]$ are considered and it
is shown that if the support is narrow then a Buchdahl inequality
holds (i.e. $2M/R<1-\epsilon,\;\epsilon>0$). This result is
superseded by the result presented here but some of the ideas in
\cite{An1} play an essential role in this work. Guven and \'{O}
Murchadha consider the general case in \cite{GM} and obtain a
bound on $2m/r$ in terms of the ratio of the tangential- and the
radial pressure, which they denote by $\gamma$. Their bound
degenerates (i.e., $2m/r\to 1$) as $\gamma\to\infty.$ (Cf. also \cite{BH}
for a similar analysis which includes a cosmological constant.) It is
interesting to note that $\gamma=\infty$ for the maximizing
solution in our work since $p=0$ and $p_T$ is a Dirac measure at
the boundary. Also note that $\gamma\to \infty$ for the sequence
constructed in \cite{An2}, which in the limit gives an infinitely
thin shell with $p=0$ and $2p_T=\rho.$ In this context we mention the work
\cite{F} where an infinitely thin shell is studied. They obtain the
bound $2M/R\leq 24/25.$ Note that this value agrees with our bound
when $\Omega=2.$ This is not surprising since their infinitely
thin shell satisfies the DEC and has $p=0$ which in our
terminology means that $\Omega=2.$ A similar study is carried out 
by Bondi \cite{Bo2} in the case $\Omega=1.$ 

Furthermore, Bondi \cite{Bo1} investigates (non-rigorously) isotropic
solutions which are allowed to have a non-monotonic energy
density. He considers models for which $\rho\geq 0,\; \rho\geq p,$
or $\rho\geq 3p,$ and obtains bounds on $2M/R$ strictly less than one in 
the respectively cases. The isotropic condition is however crucial 
since these bounds are violated for strongly non-isotropic solutions as 
this work shows (cf. also \cite{An2}, \cite{F} and \cite{Bo2}). 

The paper is organized as follows. In the next section we derive our basic inequality
which only involves $\rho$ and we formulate our main results. The main ideas of
the paper are presented in section 3. In section 4 an electrostatic analogy
(or equivalently a Newtonian analogy) is discussed and the proofs of the theorems 
are given in section 5.

\section{Set up and main results}
\setcounter{equation}{0}
Let us collect a couple of facts concerning the system (\ref{ee1})-(\ref{ee4}).
A consequence of equation (\ref{ee1}) is that
\begin{equation}\label{eminustwolambda}
e^{-2\lambda}=1-\frac{2m(r)}{r},
\end{equation}
and from (\ref{ee2}) it then follows that
\begin{equation}\label{mur}
\mu_r=(\frac{m}{r^2}+4\pi rp)e^{2\lambda}.
\end{equation}
Adding (\ref{ee1}) and (\ref{ee2}) and using the boundary conditions at $r=\infty$ gives
\begin{equation}\label{mupluslambda}
\mu(r)+\lambda(r)=-\int_r^{\infty}4\pi\eta(\rho+p)e^{2\lambda}d\eta.
\end{equation}
In particular if $R$ is the outer radius of support of the
matter then $$e^{\mu(r)+\lambda(r)}=1,$$ when $r\geq R.$ Hence,
\begin{equation}\label{muandlambda}
e^{\mu(r)}=e^{-\lambda(r)}=\sqrt{1-\frac{2m(r)}{r}},\; r\geq R.
\end{equation}
The generalized Tolman-Oppenheimer-Volkov equation (\ref{TOV2})
implies that a solution satisfies
\begin{equation}
(m+4\pi r^3p)e^{\mu+\lambda}=\int_0^r 4\pi\eta^2
e^{\mu+\lambda}(\rho+p+2p_T)d\eta.\label{fundamentaleq}
\end{equation}
Indeed, let $S=(m+4\pi r^3p)e^{\mu+\lambda}.$ Using (\ref{TOV2}) and 
(\ref{mupluslambda}) we get 
$$\frac{dS}{dr}=4\pi r^2(\rho+p+2p_T)e^{\mu+\lambda},$$ 
and the claim follows since $S(0)=0.$

Let us fix $r>0.$ Consider (\ref{fundamentaleq}), using (\ref{mupluslambda}) we get
\begin{eqnarray}
\displaystyle& & (m+4\pi r^3p)e^{-\int_{r}^{\infty}4\pi\eta(\rho+p)e^{2\lambda}d\eta}\nonumber\\
& &\displaystyle
=e^{-\int_{r}^{\infty}4\pi\sigma(\rho+p)e^{2\lambda}d\sigma}
\int_0^{r} 4\pi\eta^2
e^{-\int_{\eta}^{r}4\pi\sigma(\rho+p)e^{2\lambda}d\sigma}(\rho+p+2p_T)d\eta,
\nonumber
\end{eqnarray}
and we have
\begin{equation}
m+4\pi r^3p=\int_0^{r} 4\pi\eta^2
e^{-\int_{\eta}^{r}4\pi\sigma(\rho+p)e^{2\lambda}d\sigma}(\rho+p+2p_T)d\eta.\nonumber
\end{equation}
Since $p$ is non-negative we obtain the inequality
\begin{equation}
m(r)\leq \int_0^{r} 4\pi\eta^2
e^{-\int_{\eta}^{r}4\pi\sigma(\rho+p)e^{2\lambda}d\sigma}(\rho+p+2p_T)d\eta.\nonumber
\end{equation}
Using again the non-negativity of $p$ and the inequality (\ref{Omega}) we obtain
\begin{equation}
m(r)\leq (1+\Omega)\int_0^{r} 4\pi\eta^2\rho
e^{-\int_{\eta}^{r}4\pi\sigma\rho e^{2\lambda}d\sigma}d\eta.\label{fundamentalineq1}
\end{equation}
Note that only $\rho,$ and not $p$ and $q,$ appears in this inequality in view of
(\ref{eminustwolambda}). This is our fundamental inequality.

Let $\mathcal{B}$ be the Borel $\sigma-$algebra of $\mathbb{R}_+$
and let $\mathcal{M}$ denote the space of non-negative
$\sigma-$finite measures on $\mathcal{B}$ such that $2m(r)/r<1,$
where $m(r)=\int_{[0,r]} dh(\eta).$ Let $R>0$ and define the
operator $F_R:\mathcal{M}\to R_+$ by
\begin{equation}\label{F}
F_R(h)=\int_{[0,R]}
e^{-\int_{[r,R]}\frac{dh(\eta)}{\eta\big(1-\frac{2m(\eta)}{\eta}\big)}
}dh(r).
\end{equation}
With abuse of notation it will be understood that $F_R(u),$ where $u$ is 
a function, is the value obtained by applying $F$ to the measure $\nu$ where 
$d\nu=udr.$ Now let $\bar{\rho} =4\pi r^2\rho,$ and note that 
the inequality (\ref{fundamentalineq1}) can be written
\begin{equation}\label{ineqF}
m(r)\leq (1+\Omega)F_r(\bar{\rho}).
\end{equation}

Furthermore, note that by taking $p=0$ and $2p_T=\rho$ the inequalities
above become equalities and we can for this special class of solutions
define a form of the generalized TOV equation
which is valid whenever $4\pi r^2\rho=h\in\mathcal{M},$
\begin{equation}\label{eqF}
m(r)=(1+\Omega)F_r(h).
\end{equation}
This form of the TOV equation will be used to see that the
infinitely thin shell which maximizes $2m/r$ satisfies the TOV
equation in the sense of measures. 

By a steady state we mean a solution of the Einstein equations
(\ref{ee1})-(\ref{ee4}) such that $\rho,p$ and $p_T$ are $C^1$
functions on $[0,\infty).$ A steady state of course satisfies the
generalized Tolman-Oppenheimer-Volkov equation. For our purposes
it is sufficient that the triplet $(\rho,p,p_T)$ satisfies the
integrated form (\ref{fundamentaleq}) of the generalized TOV
equation. We say that $(\rho,p,p_T)$ is an \textit{admissible
triplet} if: each of these functions is in $L_{loc}^1([0,\infty);
4\pi r^2),$ where $4\pi r^2$ is the weight, equation
(\ref{fundamentaleq}) is satisfied a.e., and there is an
$\Omega\geq 0$ such that (\ref{Omega}) holds a.e. The following
theorem is our main result. 
\begin{theorem}
Consider any admissible triplet $(\rho,p,p_T).$ Then
\begin{equation}\label{sbound}
\sup_{r>0}\frac{2m(r)}{r}\leq\frac{(1+2\Omega)^2-1}{(1+2\Omega)^2}.
\end{equation}
\end{theorem}
The arguments leading to Theorem 2 in \cite{An1} (and also the arguments in 
the proof of Theorem 1 above) imply that the bound (\ref{sbound}) is sharp 
in the sense given by the theorem below. Before stating this theorem let us 
introduce the notation $\open{\nu}_R$ for the Dirac measure at $r=R.$

\begin{theorem}
Take $R>0,$ and let $$M=\frac{R}{2}\frac{((1+2\Omega)^2-1)}{(1+2\Omega)^2}.$$
Let $$\rho=\frac{M\open{\nu}_R}{4\pi R^2},$$ and let
$p=0$ and $2p_T=\Omega\rho,$ then (\ref{eqF}) holds with $h=4\pi
R^2\rho$ and $r=R.$
\end{theorem}

\section{Main ideas}
\setcounter{equation}{0}
The details of the proofs make the main ideas become less
transparent so let us describe them in this section.

Given a steady state with support in $[0,R],\;R>0,$ there is a
smallest $r_*\in [0,R],$ with the property that
$$\frac{2m(r_*)}{r_*}=\sup_{r>0}\frac{2m(r)}{r}.$$
We will show that if
\begin{equation}\label{if}
\frac{2m(r_*)}{r_*}>\frac{(1+2\Omega)^2-1}{(1+2\Omega)^2},
\end{equation}
then
\begin{equation}\label{then}
F_{r_*}(\bar{\rho})<\frac{m(r_*)}{1+\Omega}.
\end{equation}
In view of (\ref{ineqF}) we thus obtain a contradiction and no
steady state with the property (\ref{if}) can exist. To show that
(\ref{if}) implies (\ref{then}) is of course the main difficulty.

We will approximate the given steady state with a sum of step functions. The precise way
this is done is left to the proof.
Let $r_*$ be as above and let
\begin{equation}\label{u}
u(r)=\chi_{[r'_0,r_1]}\frac{c_1r_1}{r_1-r'_0}+\chi_{[r'_1,r_2]}\frac{c_2r_2}{r_2-r'_1}+...
+\chi_{[r'_{N-1},r_N]}\frac{c_Nr_{N}}{r_N-r'_{N-1}},
\end{equation}
where $\{r_0,r_1,...,r_N\}$ is a sub-division of the interval
$[R_0,R_1],$ so that $r_0=R_0$ and $r_N=r_*,$ and where $r_k\leq
r'_k<r_{k+1},$ and $\chi$ is the characteristic function. First we
take $r'_k=r_k$ and choose the constants $c_j,\; j=1,2,...$ so
that $u$ approximates $\bar{\rho}$ in sup norm. We will then admit
the parameters $r'_k$ to vary. Note that
$$\frac{m_u(r_k)}{r_k}=\frac{1}{r_k}\sum_{j=0}^k c_jr_j,$$ independently of the choices
of $r'_j,$ where $m_u(r):=\int_0^rudr.$

First we consider the first two terms in (\ref{u}) and
perform the limit $r'_0\to r_1$ and $r'_1\to r_2$ so that the first two step functions
become Dirac measures at $r=r_1$ and $r=r_2.$ We then show that the operator
$F$ applied to the new measure is greater than $F(u).$ More precisely we
show that
\begin{equation}\label{stratstep1}
F_{r_*}(u)<F_{r_*}(\nu_2),
\end{equation}
where
$$\nu_2=c_1r_1\open{\nu}_{r_1}+c_2r_2\open{\nu}_{r_2}+u_2 dr,$$
and
\begin{equation}\label{u2}
u_2(r)=\chi_{[r'_2,r_3]}\frac{c_3r_3}{r_3-r'_2}+\chi_{[r'_3,r_4]}\frac{c_4r_4}{r_4-r'_3}+...
+\chi_{[r'_{N-1},r_N]}\frac{c_Nr_{N}}{r_N-r'_{N-1}}.
\end{equation}
Recall that $\open{\nu}_{r_j}$ is the Dirac measure at $r=r_j.$
Clearly, a Dirac measure $\open{\nu}_{r_j},$ means that there is
an infinitely thin shell at $r=r_j$ with unit ADM mass and we will
call such a configuration a Dirac shell. The proof of
(\ref{stratstep1}) is a consequence of a crucial monotonicity
property of $F$ as $r'_0\to r_1$ and $r'_1\to r_2.$

The next step in our strategy is to show that
$$F_{r_*}(\nu_2)<F_{r_*}(\nu'_2),$$
where $\nu'_2$ is the measure obtained by \textit{moving} the Dirac shell at
$r=r_1$ to $r=r_2,$ i.e.,$$\nu'_2=(c_1r_1+c_2r_2)\open{\nu}_{r_2}+u_2dr.$$ 
It will be seen that the structure of $F$ allows one to continue this process so that
the next step is to replace the step function on the interval $[r'_2,r_3]$
by a Dirac shell with weight $c_3r_3$
at $r=r_3$ and again show that $F$ applied to this measure increases the value.
Then we move the Dirac shell with weight $c_1r_1+c_2r_2$ at $r=r_2$ to $r=r_3$
and thus obtain a Dirac shell at $r=r_3$ with weight $c_1r_1+c_2r_2+c_3r_3.$
This measure thus takes the form
$$\nu'_3=(c_1r_1+c_2r_2+c_3r_3)\open{\nu}_{r_3}+u_3dr,$$
where
\begin{equation}\nonumber
u_3(r)=\chi_{[r'_3,r_4]}\frac{c_4r_4}{r_4-r'_3}+\chi_{[r'_4,r_5]}\frac{c_5r_5}{r_5-r'_4}+...
+\chi_{[r'_{N-1},r_N]}\frac{c_Nr_{N}}{r_N-r'_{N-1}}.
\end{equation}
In this way we obtain the chain of inequalities
$$F_{r_*}(h_u)<F_{r_*}(\nu'_2)<F_{r_*}(\nu'_3)<...<F_{r_*}(\nu'_N),$$
where
\begin{equation}\label{nuprimeN}
\nu'_N=\sum_{j=1}^N c_jr_j \;\open{\nu}_{r_*}=:m_*\open{\nu}_{r_*}.
\end{equation}
Now
\begin{equation}\label{crit}
F_{r_*}(\nu'_N)=\frac{2m_*\sqrt{1-2m_*/r_*}}{1+\sqrt{1-2m_*/r_*}},
\end{equation}
which follows
by using the method in \cite{An1}, and also from the proof given in section 5.
In view of (\ref{ineqF}) we thus obtain
\begin{equation}\label{ineqshell}
m_*<\frac{2(1+\Omega)m_*\sqrt{1-2m_*/r_*}}{1+\sqrt{1-2m_*/r_*}},
\end{equation}
and solving for $2m_*/r_*$ gives
\begin{equation}\label{ineqpaper1}
\frac{2m_*}{r_*}<\frac{(1+2\Omega)^2-1}{(1+2\Omega)^2}.
\end{equation}

\section{An electrostatic analogy}
\setcounter{equation}{0}
A classical problem in electrostatics is the question
how a unit amount of charge should be spread
over a bounded set $E\in\mathbb{R}^3$ in order to minimize the Coulomb energy
$$\mathcal{E}(\rho):=\frac{1}{2}\int_E\int_E\rho(x)\rho(y)|x-y|^{2-n}dxdy.$$
Following the exposition in \cite{LL} the minimum energy is defined to be $\frac{1}{2}
\mbox{Cap}(E)^{-1},$ where $\mbox{Cap}(E)$ is the capacity of $E,$ i.e.,
\begin{equation}\label{cap}
\frac{1}{2\mbox{Cap}(E)}:=\inf\big\{\mathcal{E}(\rho):\int_{E}\rho=1\big\}.
\end{equation}
A minimizing $\rho$ does exist if $E$ is a closed set. It is not a function but a measure
(an equilibrium measure) concentrated on the surface of $E.$
In particular, if $E$ is a ball or a sphere of radius
$R$ then the optimum distribution for the charge will be
\begin{equation}\label{opt}
\rho=\frac{1}{4\pi R^2}\open{\nu}_R.
\end{equation}
and
\begin{equation}\label{capBR}
\mbox{Cap}(B_R)=R.
\end{equation}
Of course, this
problem can equivalently be formulated as a variational problem
for Newtonian gravity but since we wish to stress the relation to
capacity theory which originates from the electrostatic problem we
have preferred to use that formulation.

The analogy with our case should be clear in view of (\ref{opt}). Let us also 
note that capacity can equivalently be defined as the largest charge that can 
be carried by a body (e.g. a ball with radius $R$) if the voltage drops by 
at most one, cf. \cite{AE}. 
This formulation suggests that we in our situation define the capacity 
of a ball with radius $R$ to be the largest ADM mass that a spherically symmetric 
static body with area radius $R$ can have. 
Using this definition we then get in view of Theorem 2 that the capacity is 
given by 
$$\frac{((1+2\Omega)^2-1)R}{2(1+2\Omega)^2}.$$
Of course, we could also introduce a similar definition as in (\ref{cap}) by 
using a variational formulation for $F$ instead of $\mathcal{E}.$ 
The following theorem, taken from \cite{LL}, is an interesting feature of balls 
in $\textbf{R}^n$ for the capacity in (\ref{cap}).
\begin{theorem}(\cite{LL})
Let $E\subset \textbf{R}^n,\;n\geq 3,$ be a bounded set with
Lebesgue measure $|E|$ and let $B_E$ be the ball in $\textbf{R}^n$
with the same measure. Then $$\mbox{Cap}(B_E) \leq\mbox{Cap}(E).$$
\end{theorem}
This theorem suggests that spherical symmetry might be an important case 
also for the compactness ratio "$2M/R$" (assuming one has 
a proper definition of such a quantity) of more general static objects.

\section{Proofs}
\setcounter{equation}{0}
\textit{Proof of Theorem 1. }
Consider any admissible triplet, so that in particular $0\leq 4\pi r^2\rho\in L_{loc}^1,$
and let $f:= 4\pi r^2\rho.$ These are the only conditions of an admissible triplet
needed in this section, the remaining conditions have already been invoked to derive
the relations in section 2.
We will show that (\ref{if}) implies (\ref{then}). Hence, assume that there is
a $r_*>0$ with the property that (\ref{if}) holds. By continuity we can choose
$r_*$ so that $2m(r_*)/r_*$ is as close as we wish to the critical value and
we choose $r_*$ so that
\begin{equation}\label{rstar}
\frac{(1+2\Omega)^2-1}{(1+2\Omega)^2}<\frac{2m(r_*)}{r_*}<
\frac{1}{2}\Big(1+\frac{(1+2\Omega)^2-1}{(1+2\Omega)^2}\Big):=Q.
\end{equation}
In what follows we use the notation $m_*:=m(r_*).$ 
Fix $\epsilon >0.$
Let $\tilde{h}$ be such that $\tilde{h}=0$
on $[0,\delta)$ and $\tilde{h}=f$ on $[\delta,r_*],\;\delta>0.$ 
Obviously, for a sufficiently small $\delta>0$ the difference 
$0\leq m_f(r)-m_{\tilde{h}}(r)$ is arbitrary small and since the integration interval 
$[0,r_*]$ is finite it holds by a continuity argument that there is largest 
$\delta>0$ such that $|F_{r_*}(\tilde{h})-F_{r_*}(f)|<\epsilon/2.$ 

Now, since the operator $F$ consists of a composition of integrations, 
there is a natural number $N,$ a
sub-division $\{r_0,r_1,...,r_N\},\; r_j=\delta+j(r_*-\delta)/N,$ of
the interval $[\delta,r_*],$ and positive constants
$\{c_1,c_2,...,c_N\}$ such that the function $\bar{h}$ defined by
\begin{equation}\label{char}
\bar{h}(r)=\chi_{[r_0,r_1]}\frac{c_1r_1}{r_1-r_0}+\chi_{[r_1,r_2]}\frac{c_2r_2}{r_2-r_1}+...
+\chi_{[r_{N-1},r_N]}\frac{c_Nr_{N}}{r_N-r_{N-1}},
\end{equation}
satisfies $|F_{r_*}(\bar{h})-F_{r_*}(\tilde{h})|<\epsilon/2,$ 
and $|m_*-m_{\bar{h}}(r_*)|<\epsilon.$ 
Here $\chi_S$ is
the characteristic function, i.e., $\chi_{S}(r)=1$ if $r\in S,$ and
$\chi_{S}(r)=0$ if $r\notin S.$ The condition that the constants $c_j$ are 
positive is technical and it is not required that $f$ must be positive, 
only non-negative, but since we only seek an approximation our positivity 
condition is easy to satisfy. 
For technical reasons we also
require that $N$ is taken large, i.e.,
\begin{equation}\label{constrN}
N\geq\frac{10r_*}{(1-Q)\delta}.
\end{equation}
We now define
\begin{equation}\label{char2}
h(r)=\chi_{[r'_0,r_1]}\frac{c_1r_1}{r_1-r'_0}+\chi_{[r'_1,r_2]}\frac{c_2r_2}{r_2-r'_1}+...
+\chi_{[r'_{N-1},r_N]}\frac{c_Nr_{N}}{r_N-r'_{N-1}}.
\end{equation}
Here $r_j\leq r'_j< r_{j+1}.$ Note that $h=\bar{h}$ if $r'_j=r_j$ for all $j\in \mathbb{N}.$
Moreover note that
$$\int_0^{r_*}\bar{h}dr=\int_0^{r_*}h dr,$$ so that the quasi-local mass
at $r=r_*$ given by the energy
densities $\bar{\rho}=\bar{h}/(4\pi r^2),$ and $\rho=h/(4\pi r^2),$ are the same.
The function $h,$ will be the main object below. As explained in section 3 we will modify
$h,$ by varying the parameters $r'_j$ and moving parts of the matter, and finally obtain
the inequality
\begin{equation}\label{goal}
F_{r_*}(f)< F_{r_*}(\bar{h})+\epsilon<
F_{r_*}(\nu'_N)+\epsilon,
\end{equation}
where $\nu'_N$ is given by (\ref{nuprimeN}).
The proof is split into four steps.
\bigskip

\textit{Step 1. }\\
In the first step we will by a straightforward computation find an expression for $F_{r_*}(h).$
Since this computation is crucial and quite lengthy we will present the main steps.
In what follows $j$ and $k$ will always be non-negative integers.

Let $c_0=0,$ and let $k\geq 1.$ From (\ref{char2}) we get
\begin{equation}\label{msigma}
m(\sigma)=\sum_{j=0}^{j=k-1}c_jr_j+\frac{c_kr_k(\sigma-r'_{k-1})}{r_k-r'_{k-1}},
\mbox{ where }r_{k-1}\leq\sigma\leq r_k.
\end{equation}
By defining
\begin{equation}\nonumber
M_{k}:=\sum_{j=0}^{j=k}c_jr_j,\; k\geq 1,
\end{equation}
we thus get
\begin{equation}\label{msigma2}
m(\sigma)=M_{k-1}+\frac{c_kr_k(\sigma-r'_{k-1})}{r_k-r'_{k-1}},\,\, r_{k-1}
\leq\sigma\leq r_k.
\end{equation}
Next we define
\begin{equation}\label{Geta}
G[h](\eta)=\int_{\eta}^{\infty}
\frac{h(\sigma)\; d\sigma}{\sigma\Big(1-\frac{2m(\sigma)}{\sigma}\Big)}.
\end{equation}
Note that this is the main expression in the operator $F,$ cf. equation (\ref{F}).
From (\ref{char2}) it thus follows that for $r'_{j-1}\leq\eta\leq r_j,\; j\geq 1,$
\begin{eqnarray}
&\displaystyle G[h](\eta)=\displaystyle \int_{\eta}^{r_{j}}\frac{c_jr_j\, d\sigma}{(r_j-r'_{j-1})\sigma
\Big(1-\frac{2m(\sigma)}{\sigma}\Big)}
+\int_{r'_j}^{r_{j+1}}\frac{c_{j+1}r_{j+1}\, d\sigma}{(r_{j+1}-r'_{j})\sigma
\Big(1-\frac{2m(\sigma)}{\sigma}\Big)}&\nonumber\\
&\displaystyle +...+\int_{r'_{N-1}}^{r_{N}}\frac{c_{N}r_{N}\, d\sigma}{(r_{N}-r'_{N-1})\sigma
\Big(1-\frac{2m(\sigma)}{\sigma}\Big)}=:\tilde{G}_j(\eta)+G_{j+1}+...+G_{N}.&\nonumber
\end{eqnarray}
Here the twiddle over the first term emphasizes that it depends on $\eta$ whereas
the remaining ones do not.
By inserting the expression (\ref{msigma2}) for $m$ we get
\begin{eqnarray}\label{G1}
\tilde{G}_{j}(\eta)&=&\int_{\eta}^{r_{j}}\frac{c_jr_j\, d\sigma}{(r_j-r'_{j-1})\sigma
\Big(1-\frac{2M_{j-1}}{\sigma}-\frac{2c_jr_j(\sigma-r'_{j-1})}{\sigma(r_j-r'_{j-1})}\Big)}\nonumber\\
&=&\int_{\eta}^{r_{j}}\frac{c_jr_j\, d\sigma}
{\Big(2c_jr_jr'_{j-1}-2M_{j-1}(r_j-r'_{j-1})-\sigma(2c_jr_j-r_j-r'_{j-1})\Big)}.\nonumber
\end{eqnarray}
Note that the denominator in the integrand is positive in view of (\ref{Geta}).
Let $\Delta_j:=r_j-r'_{j-1},$ we then get
\begin{equation}\label{G1tilde}
\tilde{G}_{j}=\frac{-c_jr_j}{2c_jr_j-\Delta_j}
\log{\Big(\frac{2c_jr_jr'_{j-1}-2M_{j-1}\Delta_j -r_j(2c_jr_j-\Delta_j)}
{2c_jr_jr'_{j-1}-2M_{j-1}\Delta_j -\eta (2c_jr_j-\Delta_j)}\Big)}.
\end{equation}
Analogously we get for the $\eta$ independent terms
\begin{equation}\label{Gj}
G_{j}=\frac{-c_jr_j}{2c_jr_j-\Delta_j}
\log{\Big(\frac{2c_jr_jr'_{j-1}-2M_{j-1}\Delta_j -r_j(2c_jr_j-\Delta_j)}
{2c_jr_jr'_{j-1}-2M_{j-1}\Delta_j -r'_{j-1}(2c_jr_j-\Delta_j)}\Big)}.
\end{equation}
Let us now consider the operator $F.$ From the expression (\ref{char2}) we have
\begin{eqnarray}\label{Fh}
F_{r_*}(h)&=&\displaystyle \frac{1}{\Delta_1}\int_{r'_0}^{r_1}
c_1r_1\exp{\Big(-\tilde{G}_{1}(\eta)-\sum_{j=2}^{N}G_j\Big)}d\eta\nonumber\\
& &\displaystyle +\frac{1}{\Delta_2}\int_{r'_1}^{r_2}
c_2r_2\exp{\Big(-\tilde{G}_{2}(\eta)-\sum_{j=3}^{N}G_j\Big)}d\eta\nonumber\\
& &\displaystyle +\phantom{hej} ...\nonumber\\
& &\displaystyle+\frac{1}{\Delta_N}\int_{r'_{N-1}}^{r_N}
c_Nr_N\exp{\Big(-\tilde{G}_{N}(\eta)\Big)}d\eta.
\end{eqnarray}
Since the only dependence on $\eta$ in the integrand is in $\tilde{G}_j$ we thus obtain
\begin{eqnarray}\label{Fh1}
F_{r_*}(h)&=&\displaystyle
\frac{c_1r_1\exp{\Big(-\sum_{j=2}^{N}G_j\Big)}}{\Delta_1}\int_{r'_0}^{r_1}
\exp{\Big(-\tilde{G}_{1}(\eta)\Big)}d\eta\nonumber\\
& &\displaystyle +
\frac{c_2r_2\exp{\Big(-\sum_{j=3}^{N}G_j\Big)}}{\Delta_2}\int_{r'_1}^{r_2}
\exp{\Big(-\tilde{G}_{2}(\eta)\Big)}d\eta\nonumber\\
& &\displaystyle +\phantom{hej} ...\nonumber\\
& &\displaystyle+\frac{c_Nr_N}{\Delta_N}\int_{r'_{N-1}}^{r_N}
\exp{\Big(-\tilde{G}_{N}(\eta)\Big)}d\eta.
\end{eqnarray}
The first two terms in this expression can be written as
\begin{equation}\label{G1G2}
\Big(\frac{c_1r_1 e^{-G_2}}{\Delta_1}\int_{r'_0}^{r_1}
e^{-\tilde{G}_{1}(\eta)}d\eta+
\frac{c_2r_2}{\Delta_2}\int_{r'_1}^{r_2}
e^{-\tilde{G}_{2}(\eta)}d\eta\Big)e^{-\sum_{j=3}^{N}G_j}.
\end{equation}
As explained in section 3 the idea is to show that 
$F_{r_*}(h)$ is dominated by $F_{r_*}(\nu_2),$ where
$\nu_2$ is the measure
\begin{equation}\label{nu2}
\nu_2(r)=c_1r_1\open{\nu}_{r_1}+c_2r_2\open{\nu}_{r_2}+\chi_{[r'_2,r_3]}\frac{c_3r_3}{r_3-r'_2}
+...+\chi_{[r'_{N-1},r_N]}\frac{c_Nr_{N}}{r_N-r'_{N-1}},
\end{equation}
and then to show that $F_{r_*}(\nu_2)<F_{r_*}(\nu'_2)$ where
\begin{equation}\label{nuprime2}
\nu'_2(r)=(c_1r_1+c_2r_2)\open{\nu}_{r_2}+\chi_{[r'_2,r_3]}\frac{c_3r_3}{r_3-r'_2}
+...+\chi_{[r'_{N-1},r_N]}\frac{c_Nr_{N}}{r_N-r'_{N-1}}.
\end{equation}
The measure $\nu'_2$ can thus be thought of as a modified $h$ where $c_1$
and $c_2$ have been replaced by $c'_1=0$ and $c'_2=(c_1r_1+c_2r_2)/r_2$ respectively,
and where the limit $r'_1\to r_2$ has been carried out. Note that the quasi-local mass
generated by $\nu'_2$ and $h$ are the same, i.e., $m_{\nu'_2}(r_*)=m_h(r_*).$ In order to 
show that $F_{r_*}(h)< F_{r_*}(\nu'_2),$ the terms in the bracket in (\ref{G1G2}) must 
be dominated by
\begin{equation}\label{G1G22}
\Big(\lim_{r'_0\to r_1}\frac{c_1r_1 e^{-G_2}}{\Delta_1}\int_{r'_0}^{r_1}
e^{-\tilde{G}_{1}(\eta)}d\eta+
\lim_{r'_1\to r_2}\frac{c_2r_2}{\Delta_2}\int_{r'_1}^{r_2}
e^{-\tilde{G}_{2}(\eta)}d\eta\Big),
\end{equation}
which in turn must be dominated by 
\begin{equation}\label{Gprim2}
\lim_{r'_1\to r_2}\frac{c'_2r_2}{\Delta_2}\int_{r'_1}^{r_2}
e^{-\tilde{G'}_{2}(\eta)}d\eta.
\end{equation}
Here $\tilde{G'}_2$ denotes the $G-$function which corresponds to
the measure $\nu'_2.$ The structure of $F(h)$ revealed in (\ref{Fh1}) then shows
that this procedure can be continued: we define the measures $\nu_3$ and 
$\nu'_3$ by
\begin{equation}\nonumber
\nu_3(r)=(c_1r_1+c_2r_2)\open{\nu}_{r_2}+c_3r_3\open{\nu}_{r_3}
+\chi_{[r'_3,r_4]}\frac{c_4r_4}{r_4-r'_3}
+...+\chi_{[r'_{N-1},r_N]}\frac{c_Nr_{N}}{r_N-r'_{N-1}},
\end{equation}
\begin{equation}\nonumber
\nu'_3(r)=(c_1r_1+c_2r_2+c_3r_3)\open{\nu}_{r_3}+\chi_{[r'_3,r_4]}\frac{c_4r_4}{r_4-r'_3}
+...+\chi_{[r'_{N-1},r_N]}\frac{c_Nr_{N}}{r_N-r'_{N-1}},
\end{equation}
and we show that $F_{r_*}(\nu'_2)<F_{r_*}(\nu_3)<F_{r_*}(\nu'_3).$
In this way we obtain a chain of inequalities
$$F_{r_*}(\bar{h})< F_{r_*}(\nu'_2)< F_{r_*}(\nu'_3)< ...< F_{r_*}(\nu'_N),$$ where $\nu'_N$ is
the Dirac measure at $r=r_N=r_*$ with $m_{\nu'_N}(r_*)=m_{\bar{h}}(r_*).$
Let us now compute the sum of the two terms in the bracket in (\ref{G1G2}).
We use the following notation
\begin{equation}\label{T1}
T_1=\frac{c_1r_1 e^{-G_2}}{\Delta_1}\int_{r'_0}^{r_1}
e^{-\tilde{G}_{1}(\eta)}d\eta,
\end{equation}
and
\begin{equation}\label{T2}
T_2=\frac{c_2r_2}{\Delta_2}\int_{r'_1}^{r_2}
e^{-\tilde{G}_{2}(\eta)}d\eta.
\end{equation}
We have from (\ref{G1tilde})
\begin{eqnarray}\label{T1comp1}
\displaystyle\int_{r'_0}^{r_1}e^{-\tilde{G}_{1}(\eta)}d\eta&=&\int_{r'_0}^{r_1}
\Big(\frac{2c_1r_1r'_0-r_1(2c_1r_1-\Delta_1)}{2c_1r_1r'_0-\eta(2c_1r_1-\Delta_1)}\Big)^
{\frac{c_1r_1}{2c_1r_1-\Delta_1}}d\eta\nonumber\\
&=&\displaystyle \frac{\Big(2c_1r_1r'_0-r_1(2c_1r_1-\Delta_1)\Big)^{\frac{c_1r_1}{2c_1r_1-\Delta_1}}}
{\Big(1-\frac{c_1r_1}{2c_1r_1-\Delta_1}\Big)(2c_1r_1-\Delta_1)}\nonumber\\
& &\displaystyle\times\Big[-\Big(2c_1r_1r'_0-\eta(2c_1r_1-\Delta_1)\Big)^{1-\frac{c_1r_1}{2c_1r_1-\Delta_1}}\Big]
_{r'_0}^{r_1}\nonumber\\
&=&\displaystyle \frac{\Big(2c_1r_1r'_0-r_1(2c_1r_1-\Delta_1)\Big)^{\frac{c_1r_1}{2c_1r_1-\Delta_1}}}
{c_1r_1-\Delta_1}\nonumber\\
& &\displaystyle\times\Big\{\Big(2c_1r_1r'_0-r'_0(2c_1r_1-\Delta_1)\Big)^
{1-\frac{c_1r_1}{2c_1r_1-\Delta_1}}\nonumber\\
& &-\Big(2c_1r_1r'_0-r_1(2c_1r_1-\Delta_1)\Big)^
{1-\frac{c_1r_1}{2c_1r_1-\Delta_1}}\Big\}\nonumber\\
&=&\displaystyle \frac{\Big(2c_1r_1r'_0-r_1(2c_1r_1-\Delta_1)\Big)}{c_1r_1-\Delta_1}
\nonumber\\
& &\displaystyle\times\Big\{\Big(\frac{2c_1r_1r'_0-r'_0(2c_1r_1-\Delta_1)}{2c_1r_1r'_0-r_1(2c_1r_1-\Delta_1)}\Big)^{{\frac{c_1r_1-\Delta_1}{2c_1r_1-\Delta_1}}}-1\Big\}.
\end{eqnarray}
Furthermore, from (\ref{Gj}) we have
\begin{equation}\nonumber
e^{-G_2}= \Big(\frac{2c_2r_2r'_1-2M_{1}\Delta_2-r_2(2c_2r_2-\Delta_2)}
{2c_2r_2r'_1-2M_{1}\Delta_2-r'_1(2c_2r_2-\Delta_2)}\Big)^{\frac{c_2r_2}{2c_2r_2-\Delta_2}}.
\end{equation}
The term $T_1$ can thus be written
\begin{eqnarray}\label{T1result}
T_1&=&\frac{c_1r_1}{\Delta_1}\frac{\Big(2c_1r_1r'_0-r_1(2c_1r_1-\Delta_1)\Big)}{c_1r_1-\Delta_1}\nonumber\\
& &\times\Big\{\Big(\frac{2c_1r_1r'_0-r'_0(2c_1r_1-\Delta_1)}{2c_1r_1r'_0-r_1(2c_1r_1-\Delta_1)}\Big)^{\frac{c_1r_1-\Delta_1}{2c_1r_1-\Delta_1}}-1\Big\}\nonumber\\
& &\times\Big(\frac{2c_2r_2r'_1-2M_{1}\Delta_2-r_2(2c_2r_2-\Delta_2)}
{2c_2r_2r'_1-2M_{1}\Delta_2-r'_1(2c_2r_2-\Delta_2)}\Big)^{\frac{c_2r_2}{2c_2r_2-\Delta_2}}.
\end{eqnarray}
A very similar calculation shows that
\begin{eqnarray}\label{T2result}
T_2&=&\frac{c_2r_2}{\Delta_2}\frac{\Big(2c_2r_2r'_1-2M_{1}\Delta_2-r_2(2c_2r_2-\Delta_2)\Big)}{c_2r_2-\Delta_2}\nonumber\\
& &\times
\Big\{\Big(\frac{2c_2r_2r'_1-2M_{1}\Delta_2-r'_1(2c_2r_2-\Delta_2)}
{2c_2r_2r'_1-2M_{1}\Delta_2-r_2(2c_2r_2-\Delta_2)}\Big)^{\frac{c_2r_2-\Delta_2}{2c_2r_2-\Delta_2}}-1\Big\}.
\end{eqnarray}
The aim is to obtain the inequality $F_{r_*}(\bar{h})\leq F_{r_*}(\nu'_2).$ Since
$h$ and $\nu'_2$ are identical for $r\geq r_3$ it follows from (\ref{Fh1}) that it
is sufficient to
obtain the estimate $T_1+T_2\leq T_1^{\nu'_2}+T_2^{\nu'_2},$ where
$T_1^{\nu'_2}$ and $T_2^{\nu'_2}$ are the corresponding terms for $\nu'_2.$ Clearly
$T_1^{\nu'_2}=0$ since $c'_1=0,$ and $T_2^{\nu'_2}$ is the expression (\ref{Gprim2}) which
in view of (\ref{T2result}) and the fact that $M_1=0$ in this case since $c'_1=0$
is given by
\begin{equation}\label{Tprim2}
T_2^{\nu'_2}=\lim_{r'_1\to r_2}T'_2,
\end{equation}
where
\begin{eqnarray}\label{Tnu2result}
T'_2&=&\frac{c'_2r_2}{\Delta_2}\frac{\Big(2c'_2r_2r'_1-r_2(2c'_2r_2-\Delta_2)\Big)}{c'_2r_2-\Delta_2}\nonumber\\
& &\times
\Big\{\Big(\frac{2c'_2r_2r'_1-r'_1(2c'_2r_2-\Delta_2)}
{2c'_2r_2r'_1-r_2(2c'_2r_2-\Delta_2)}\Big)^{\frac{c'_2r_2-\Delta_2}{2c'_2r_2-\Delta_2}}-1\Big\}.
\end{eqnarray}
Here $c'_2=(c_1r_1+c_2r_2)/r_2.$
The expressions for $T_1,\,T_2$ and $T'_2$ will now be simplified. Let us introduce
the notation
$$b_k=\frac{r'_{k-1}}{r_k},\; k=1,2,...$$
which implies that
$$\Delta_k=r_k(1-b_k).$$
Let us consider the term $T_2.$ By dividing both the numerator and the denominator
by $2c_2r_{2}^{2},$ the first factor in the expression (\ref{T2result})
can be written
\begin{eqnarray}\label{T2simpcomp}
& &\frac{c_2r_2}{\Delta_2}\frac{\Big(2c_2r_2r'_1-2M_{1}\Delta_2-r_2(2c_2r_2-\Delta_2)\Big)}{c_2r_2-\Delta_2}\nonumber\\
& & \displaystyle =\frac{c_2r_2}{1-b_2}
\frac{\Big(b_2-\frac{M_1(1-b_2)}{c_2r_2}-(1-\frac{1-b_2}{2c_2})\Big)}{\frac{c_2r_2}{2c_2r_2}-\frac{1-b_2}{2c_2}}
\nonumber\\
& &=\frac{c_2r_2}{1-b_2}
\frac{\Big((1-b_2)(\frac{1}{2c_2}-\frac{c_1r_1}{c_2r_2} -1)\Big)}{\frac{1}{2}-\frac{1-b_2}{2c_2}}
=\frac{c_2r_2(1-2c_1\frac{r_1}{r_2}-2c_2)}{c_2-(1-b_2)}.
\end{eqnarray}
The second factor can be simplified in a similar way
\begin{eqnarray}\label{T2simpcomp2}
& &\Big(\frac{2c_2r_2r'_1-2M_{1}\Delta_2-r'_1(2c_2r_2-\Delta_2)}
{2c_2r_2r'_1-2M_{1}\Delta_2-r_2(2c_2r_2-\Delta_2)}\Big)^{\frac{c_2-(1-b_2)}{2c_2-(1-b_2)}}-1\nonumber\\
& &=\Big(\frac{b_2-\frac{c_{1}r_1}{c_2r_2}(1-b_2)-b_2(1-\frac{1-b_2}{2c_2})}
{b_2-\frac{c_{1}r_1}{c_2r_2}(1-b_2)-(1-\frac{1-b_2}{2c_2})}\Big)^{\frac{c_2-(1-b_2)}{2c_2-(1-b_2)}}-1\nonumber\\
& &=\Big(\frac{(\frac{b_2}{2c_2}-\frac{c_{1}r_1}{c_2r_2})(1-b_2)}{(1-b_2)(\frac{1}{2c_2}
-\frac{c_1r_1}{c_2r_2})}\Big)^{\frac{c_2-(1-b_2)}{2c_2-(1-b_2)}}-1\nonumber\\
& &=\Big(\frac{b_2-2c_1\frac{r_1}{r_2}}{1-2c_1\frac{r_1}{r_2}-2c_2}\Big)^{\frac{c_2-(1-b_2)}{2c_2-(1-b_2)}}-1.
\end{eqnarray}
In conclusion $T_2$ can be written
\begin{equation}\label{T2simpresult}
T_2=\frac{c_2r_2(1-2c_1\frac{r_1}{r_2}-2c_2)}{c_2-z_2}\Big\{\Big(\frac{1-2c_1\frac{r_1}{r_2}-z_2}{1-2c_1\frac{r_1}{r_2}-2c_2}\Big)^{\frac{c_2-z_2}{2c_2-z_2}}-1\Big\},
\end{equation}
where we have introduced the notation
$$z_k=1-b_k,\;k=1,2,...$$
Simplifying $T_1$ and $T'_2$ in a similar way leads to the following expressions
\begin{equation}\label{T1simpresult}
T_1=\frac{c_1r_1(1-2c_1)}{c_1-z_1}\Big\{\Big(\frac{1-z_1}{1-2c_1}\Big)^{\frac{c_1-z_1}{2c_1-z_1}}-1\Big\}\Big(\frac{1-2c_1\frac{r_1}{r_2}-2c_2}{1-2c_1\frac{r_1}{r_2}-z_2}\Big)
^{\frac{c_2}{2c_2-z_2}},
\end{equation}
and
\begin{equation}\label{T2primsimpresult}
T'_2=\frac{(c_1r_1+c_2r_2)(1-2c_1\frac{r_1}{r_2}-2c_2)}{c_2+c_1\frac{r_1}{r_2}-z_2}\Big\{\Big(\frac{1-z_2}{1-2c_1\frac{r_1}{r_2}-2c_2}\Big)^{\frac{c_2+c_1\frac{r_1}{r_2}-z_2}{2c_2+2c_1\frac{r_1}{r_2}-z_2}}-1\Big\}.
\end{equation}
Note that the expression for $T'_2$ is obtained from (\ref{T2simpresult}) by putting $c_1=0$ and replacing $c_2$ by $c_2+c_1r_1/r_2$ in accordance with the previous discussion.

\bigskip

\textit{Step 2.}\\
In this step we show that $F_{r_*}(\bar{h})<F_{r_*}(\nu_2)$ by showing that $F$ is monotone 
as $r'_0\to r_1$ and
$r'_1\to r_2,$ i.e., as $z_1\to 0$ and $z_2\to 0.$
Let us define
\begin{equation}\label{A}
A(z,c)=\frac{1}{c-z}\Big\{\Big(\frac{1-z}{1-2c}\Big)^{\frac{c-z}{2c-z}}-1\Big\},
\end{equation}
and
\begin{equation}\label{B}
B(z,c)=\Big(\frac{1-2c}{1-z}\Big)^{\frac{c}{2c-z}},
\end{equation}
where $z\in [0,1/10],$ and $c\in (0,Q/2).$ Recall the definition of $Q$ in (\ref{rstar}).
These are the fundamental functions in the expressions for $T_1$ and $T_2,$ namely
\begin{equation}\label{T1AB}
T_1=c_1r_1(1-2c_1)A(z_1,c_1)B\Big(\frac{z_2}{1-2c_1r_1/r_2},\frac{c_2}{1-2c_1r_1/r_2}\Big),
\end{equation}
and
\begin{equation}\label{T2AB}
T_2=\frac{c_2r_2(1-2c_1\frac{r_1}{r_2}-2c_2)}{1-2c_1r_1/r_2}A\Big(\frac{z_2}{1-2c_1r_1/r_2},\frac{c_2}{1-2c_1r_1/r_2}\Big).
\end{equation}
Let us now see that the domain of definition of the functions $A$
and $B$ is relevant. Since $r_*$ is the smallest $r$ with $2m_*/r_*=Q$ we have in view 
of (\ref{rstar}) $c_1r_1+c_2r_2<r_2Q/2,$ and since $Q<1,$ $Qc_1r_1+c_2r_2<r_2Q/2,$
which implies that
$$c_2r_2<\frac{Q}{2}r_2(1-2c_1r_1/r_2),$$
and thus
$$c_2/(1-2c_1r_1/r_2)<Q/2.$$
Since $c_1<Q/2$ it follows that the second argument in $A$ and $B$
in (\ref{T1AB}) and (\ref{T2AB}) is less than $Q/2,$ i.e., $c\in
(0,Q/2).$ To see that the first argument in the functions $A$ and
$B$ belong to $[0,1/10]$ we first check that the condition
(\ref{constrN}) implies that
$$\frac{\delta+(r_*-\delta)/N}{\delta+2(r_*-\delta)/N}\geq\frac{9+Q}{10}.$$
The inequality above can be written
$$\frac{(8+2Q)(r_*-\delta)}{10N}\leq\frac{(1-Q)\delta}{10},$$
which clearly is satisfied if $$N\geq \frac{10r_*}{(1-Q)\delta}.$$
In view of (\ref{constrN}) we thus have for $j\geq 1,$ 
\begin{equation}\label{quotients}
\frac{r'_j}{r_{j+1}}\geq\frac{r_j}{r_{j+1}}\geq\frac{r_1}{r_2}
=\frac{\delta+(r_*-\delta)/N}{\delta+2(r_*-\delta)/N}\geq\frac{9+Q}{10},
\end{equation}
so that $z_k=1-r'_{k-1}/r_k\leq (1-Q)/10$ for all $k.$
It follows that
$$\frac{z_2}{1-2c_1r_1/r_2}<\frac{z_2}{1-Q}\leq\frac{1}{10},$$
which proves our claim, i.e., $z\in [0,1/10].$


It is clear that these facts hold in general, i.e., not only for the terms
$T_1$ and $T_2$ but at any step in our chain of inequalities since
$\sum_{j=1}^k c_jr_j<r_kQ/2.$ We can of course also express the term $T'_2$ in a similar way but it is not useful here. By construction the functions $A$ and $B$ are continuous in
the domain of definition,
in particular they are continuous along the lines $z=c$ and $z=2c.$

\begin{lemma}
For any $c\in (0,Q/2)$ the functions $A(\cdot,c)$ and $B(\cdot,c)$
are decreasing in $z,\;z\in [0,1/10].$
\end{lemma}
\textit{Proof of Lemma 1. } Monotonicity of $A.$ Let us introduce
the new variables
\begin{equation}\label{betaandk}
\beta=\frac{2c-z}{c}\mbox{ and }k=\frac{c}{1-2c}.
\end{equation}
We thus have that $$0<k\leq \frac{Q}{2(1-Q)}, \mbox{ and
}\beta\leq 2.$$ We now express $A$ in terms of these variables and by
abuse of notation we denote this function again by $A.$ Since
$$\frac{1-z}{1-2c}=1+\frac{2c-z}{1-2c}=1+k\beta,$$
it follows that
\begin{equation}\label{Abetak}
A(\beta,k)=\frac{1+2k}{k}\Big\{(1+k\beta)^{\frac{\beta-1}{\beta}}-1\Big\}.
\end{equation}
We now want to show that $\partial_{\beta}A\geq 0$ since $\partial_z\beta$ is negative.
A straightforward computation gives after some rearrangements
\begin{eqnarray}\label{partialA}
\partial_{\beta}A&=&\frac{(1+k\beta)^{\frac{\beta-1}{\beta}}}{(1-\beta)^2\beta^2(1+k\beta)}
\Big[-\beta^2(1+k\beta)+\beta^2(1+k\beta)^{\frac{1}{\beta}}\nonumber\\
& &+(\beta-1)(1+k\beta)\log{(1+k\beta)}+(1-\beta)^2k\beta\Big].
\end{eqnarray}
Let us denote the factor in square brackets by $\Psi.$ Adding the first and the last
term in this expression gives
\begin{equation}\label{Psi}
\Psi(\beta,k)=\beta^2(1+k\beta)^{\frac{1}{\beta}}+(\beta-1)(1+k\beta)\log{(1+k\beta)}-\beta^2-k\beta(2\beta-1).
\end{equation}
Let
\begin{equation}\label{gamma}
\gamma=\frac{\log{(1+k\beta)}}{\beta},
\end{equation}
which is well defined also when $\beta=0$ since $\lim_{\beta\to
0}\gamma=k.$ Since $$k\beta=\frac{2c-z}{1-2c},$$ it follows that
$k\beta<1/(1-Q),$ and since $k\beta$ is positive as long as
$2c\geq 1/10$ a rough estimate gives
\begin{equation}\label{kbeta}
k\beta\geq -1/10, 
\end{equation}
by the condition that $z\leq 1/10.$ We will below distinguish between
the two cases $0\leq \beta\leq 2,$ and $\beta<0.$ In both cases
$\gamma >0,$ or more precisely, in the former case we have
$\gamma\in [\log{(1+2k)}/2,k],$ and in the latter case $\gamma\in
(0,k].$ By using the relation $$k\beta=e^{\gamma\beta}-1,$$ $\Psi$
takes the form
\begin{equation}\label{Psigammak}
\Psi(\beta,\gamma)=-1+2\beta+\beta^2(e^{\gamma}-1)+e^{\gamma\beta}[(\beta-1)\beta\gamma
-2\beta+1].
\end{equation}
By expanding the exponential functions using the formula
$e^x=1+x/1!+x^2/2!+...$ and collecting the terms corresponding to
different powers in $\gamma$ gives
\begin{equation}\label{Psiexpand}
\Psi(\beta,\gamma)=\beta^2\sum_{j=3}^{\infty}\Big[\frac{1}{j!}-\beta^{j-2}\Big(\frac{1}{(j-1)!}-\frac{1}{j!}\Big)
+\beta^{j-1}\Big(\frac{1}{(j-1)!}-\frac{2}{j!}\Big)\Big]\gamma^j.
\end{equation}
Note that the lower orders of $\gamma$ vanish.
We denote the factors in square brackets by $\Phi_j,$ and these can thus be written as
\begin{equation}\label{Phi}
\Phi_j(\beta)=\frac{1}{j!}\Big(1-(j-1)\beta^{j-2}+(j-2)\beta^{j-1}\Big).
\end{equation}
We now claim that
\begin{equation}\label{Psiclaim}
\Phi_j(\beta)=\frac{(1-\beta)^2}{j!}(1+2\beta+3\beta^2+...+(j-2)\beta^{j-3}),\; j\geq 3.
\end{equation}
This statement is easily shown by an induction argument. First, if $j=3$ we have from
(\ref{Phi}) that
\[
\Phi_3=\frac{1}{3!}(1-2\beta+\beta^2)=\frac{1}{3!}(1-\beta)^2,
\]
so the claim is true for $j=3.$
Assume now that for any positive integer $P\geq 2,$
\begin{equation}\label{indhyp}
(1-\beta^P(P+1)+\beta^{P+1}P)=(1-\beta^2)(1+2\beta+3\beta^2+...+P\beta^{P-1}).
\end{equation}
We then have by (\ref{indhyp})
\begin{eqnarray*}
& &(1-\beta^{P+1}(P+2)+\beta^{P+2}(P+1))\nonumber\\
& &=(1-\beta^2)(1+2\beta+3\beta^2+...+P\beta^{P-1})
\nonumber\\
& &\;\;\;\;+\beta^P(P+1)-2(P+1)\beta^{P+1}+\beta^{P+2}(P+1)\nonumber\\
& &=(1-\beta^2)(1+2\beta+3\beta^2+...+P\beta^{P-1})+\beta^P(P+1)(1-\beta)^2\nonumber\\
& &=(1-\beta^2)(1+2\beta+3\beta^2+...+(P+1)\beta^{P}),
\end{eqnarray*}
and the claim (\ref{Psiclaim}) follows.
In conclusion we have shown
\begin{equation}\label{Aprimfinalf}
\partial_{\beta}A=(1+k\beta)^{\frac{-1}{\beta}}
\sum_{j=3}^{\infty}\frac{1}{j!}[1+2\beta+3\beta^2+...+(j-2)\beta^{j-3}]\gamma^{j}.
\end{equation}
Note that the lower orders of $\gamma$ have vanished. 
Now, $1+k\beta >0,$ since $k\beta\geq -1/10,$ and $\gamma> 0,$ so in
the case $\beta\geq 0,$ it follows immediately that
$\partial_{\beta}A\geq 0.$ Let us therefore consider the remaining
case $\beta<0.$ First we note that $\beta<0$ implies that $z>2c.$
Now, since $z\leq 1/10$ this means that $\beta$ is only negative if
$c$ is small, i.e., $c<1/20.$ Therefore, since $\gamma\leq k$ we get 
\begin{equation}\label{gammaest}
\gamma\leq k=\frac{c}{1-2c}<1/18.
\end{equation}
From the inequality (cf. \cite{AS})
$$|\log{(1-x})|<\frac{3x}{2},\;\;\; 0<x\leq 1/2,$$
we have
\begin{equation}\label{gammabeta}
|\gamma\beta|=|\log{(1-k|\beta|)}|<3k|\beta|/2\leq 3/20,
\end{equation}
where the last inequality followed from (\ref{kbeta}). Let us now
estimate the sum in (\ref{Aprimfinalf}). For this we use that
\[
\frac{1}{4!}+\frac{\gamma}{5!}+\frac{\gamma^2}{6!}+...<\frac{1}{4!}
(1+\gamma+\gamma^2+...)=\frac{1}{4!}\frac{1}{(1-\gamma)}<\frac{1}{20},
\]
by (\ref{gammaest}), together with the formula
\[
1+2x+3x^2+...=\frac{1}{(1-x)^2},\;\;\; -1<x<1.
\]
We drop the non-negative terms except for the first one and obtain
\begin{eqnarray}\label{sumest}
& &\frac{1}{3!}+\frac{1}{4!}(1+2\beta)\gamma+\frac{1}{5!}(1+2\beta+3\beta^2)\gamma^2+
\frac{1}{6!}(1+2\beta+3\beta^2+4\beta^3)\gamma^3+...
\nonumber\\
& &\geq\frac{1}{3!}-\Big[2\gamma|\beta|\Big(\frac{1}{4!}+\frac{\gamma}{5!}+\frac{\gamma^2}{6!}\Big)
+4(\gamma|\beta|)^3\Big(\frac{1}{6!}+\frac{\gamma}{7!}+\frac{\gamma^2}{8!}+...\Big)+...\Big]
\nonumber\\
& &\geq\frac{1}{3!}-\frac{1}{20}\Big[2\gamma|\beta|+4(\gamma|\beta|)^3+6(\gamma|\beta|)^5...]\nonumber\\
& &\geq\frac{1}{3!}-\frac{1}{20}\Big[1+2\gamma|\beta|+3(\gamma|\beta|)^2+4(\gamma|\beta|)^3+...]\nonumber\\
& &=\frac{1}{3!}
-\frac{1}{20(1-\gamma|\beta|)^2}\geq\frac{1}{3!}-\frac{20^2}{20\cdot 17^2}>0.
\end{eqnarray}
In the second last inequality we used (\ref{gammabeta}) and in the
last (\ref{kbeta}). Thus $\partial_{\beta}A>0$ also in the case
when $\beta<0,$ and the monotonicity of $A(\cdot,c)$ follows. Let
us now turn to the
monotonicity of $B(\cdot,c).$\\

Monotonicity of $B$. We express $B$ in the variables $k$ and
$\beta$ and, by abuse of notation, get
\[
B(\beta,k)=(1+k\beta)^{\frac{-1}{\beta}}.
\]
As in the case of the function $A$ the claimed monotonicity follows if we can show
\[
\partial_{\beta}B\geq 0.
\]
We have
\[
\partial_{\beta}B=\frac{(1+k\beta)^{\frac{-1}{\beta}}}{\beta^2(1+k\beta)}[(1+k\beta)\log{(1+k\beta)}-k\beta].
\]
Using the variable $\gamma$ defined in (\ref{gamma}) we can write the factor in square
brackets as
\[
[(1+k\beta)\log{(1+k\beta)}-k\beta]=\gamma\beta e^{\gamma\beta}-(e^{\gamma\beta}-1).
\]
By letting $a=\gamma\beta$ we have a function of one variable and it is elementary to
show the non-negativity of this expression for any $a.$
This completes the proof of the lemma.
\begin{flushright}
$\Box$
\end{flushright}

\textit{Step 3.}\\
In this step we show that $F(\nu_2)<F(\nu'_2).$ 
Hence, we want to show that 
$$0\leq \lim_{z_2\to 0}T'_2-\lim_{z_2\to 0}\Big(\lim_{z_1\to 0}T_1\Big)-\lim_{z_2\to 0}T_2
=:\bar{T}'_2-\bar{T}_1-\bar{T}_2.$$ We have from
(\ref{T2simpresult})-(\ref{T2primsimpresult})
\[
\bar{T}_1=r_1\Big(\sqrt{\frac{1-2c_1}{1-2c_1r_1/r_2}}-\frac{1-2c_1}{\sqrt{1-2c_1r_1/r_2}}\Big)\sqrt{1-2c'_2},
\]
\[
\bar{T}_2=r_2\Big(\sqrt{1-2c_1r_1/r_2}-\sqrt{1-2c'_2}\Big)\sqrt{1-2c'_2},
\]
and
\[
\bar{T}'_2=r_2\Big(1-\sqrt{1-2c'_2}\Big)\sqrt{1-2c'_2}.
\]
Hence
\begin{eqnarray}\nonumber
\bar{T}'_2-\bar{T}_1-\bar{T}_2&=&\sqrt{1-2c'_2}\Big[r_2(1-\sqrt{1-2c_1r_1/r_2})
\nonumber\\
& &-r_1\sqrt{\frac{1-2c_1}{1-2c_1r_1/r_2}}(1-\sqrt{1-2c_1})\Big].
\end{eqnarray}
Define $$\kappa:=1-\frac{r_1}{r_2},\mbox{ so that }\kappa\in (0,1),$$ then
$$\sqrt{1-2c_1r_1/r_2}=\sqrt{1-2c_1}\sqrt{1+\frac{2c_1\kappa}{1-2c_1}}.$$
The factor in square brackets above can be written
\begin{eqnarray}\nonumber
\Gamma:&=&r_2\Big(1-\sqrt{1-2c_1\frac{r_1}{r_2}}\Big)
-r_1\sqrt{\frac{1-2c_1}{1-2c_1\frac{r_1}{r_2}}}(1-\sqrt{1-2c_1})\nonumber\\
&=&r_2\Big(1-\sqrt{1-2c_1}\sqrt{1+\frac{2c_1\kappa}{1-2c_1}}\Big)\nonumber\\
& &
-r_1\frac{1}{\sqrt{1+\frac{2c_1\kappa}{1-2c_1}}}(1-\sqrt{1-2c_1})\nonumber\\
&=&\frac{r_2}{\sqrt{1+\frac{2c_1\kappa}{1-2c_1}}}
\Big[\sqrt{1+\frac{2c_1\kappa}{1-2c_1}}
-\sqrt{1-2c_1}\Big(1+\frac{2c_1\kappa}{1-2c_1}\Big)\nonumber\\
& &-(1-\kappa)(1-\sqrt{1-2c_1})\Big]\nonumber\\
&=&\frac{r_2}{\sqrt{1+\frac{2c_1\kappa}{1-2c_1}}}
\Big[\sqrt{1+\frac{2c_1\kappa}{1-2c_1}}-\frac{\kappa}{\sqrt{1-2c_1}}+\kappa-1\Big]
\nonumber\\
&=&\frac{r_2}{\sqrt{1-2c_1(1-\kappa)}}
\Big[\sqrt{1-2c_1(1-\kappa)}-\kappa-\sqrt{1-2c_1}(1-\kappa))\Big].\nonumber
\end{eqnarray}
Let us introduce the notation
\begin{equation}\label{Gamma}
\Gamma(c_1,\kappa):=\sqrt{1-2c_1(1-\kappa)}-\kappa-\sqrt{1-2c_1}(1-\kappa)),
\end{equation}
so that
\[
\bar{T}'_2-\bar{T_1}-\bar{T_2}=r_2\sqrt{\frac{1-2c'_2}{1-2c_1(1-\kappa)}}
\Gamma(c_1,\kappa).
\]
We want to show that the right hand side is non-negative for any admitted choice
of the parameters $c_1, c_2$ and $\kappa.$
Since $\Gamma(0,\kappa)=0$ the statement follows since $\partial_{c_1}\Gamma>0.$
Indeed, we have
\[
\frac{\partial\Gamma}{\partial c_1}=(1-\kappa)\Big[\frac{1}{\sqrt{1-2c_1}}-\frac{1}{\sqrt{1-2c_1(1-\kappa)}}\Big],
\]
which is positive since $\kappa\in (0,1).$ Hence $\bar{T}'_2-\bar{T}_1-\bar{T}_2>0.$\\ 

\bigskip
\textit{Step 4.}\\
At this stage it is clear that by repeating the arguments we obtain 
\begin{equation}\nonumber
F_{r_*}(\bar{h})<F_{r_*}(\nu'_2)<F_{r_*}(\nu'_3)<...<F_{r_*}(\nu'_N), 
\end{equation}
where $\nu'_N$ is the Dirac measure at $r=r_N=r_*$ 
with $m_{\nu'_N}(r_*)=m_{h}(r_*).$ 
An appropriate method for computing $F_{r_*}(\nu'_N)$ is given in \cite{An1}. 
However, we can also use the formula (\ref{T2primsimpresult}) with $c_1=0,$  
$c_2=m_h(r_*)/r_*$ and $z_2=0,$ and we get with $m'_*:=m_h(r_*)$
\begin{equation}\label{FDs}
F_{r_*}(\nu'_N)=r_*(1-\frac{2m'_*}{r_*})\Big\{\frac{1}{\sqrt{1-\frac{2m'_*}{r_*}}}-1\Big\}
=\frac{2m'_*\sqrt{1-\frac{2m'_*}{r_*}}}{1+\sqrt{1-\frac{2m'_*}{r_*}}}.
\end{equation}
The inequalities (\ref{ineqF}) and (\ref{goal}) then gives 
\begin{equation}\label{ineqshell}
m_*<\frac{2(1+\Omega)m'_*\sqrt{1-2m'_*/r_*}}{1+\sqrt{1-2m'_*/r_*}}+\epsilon,
\end{equation}
Using that $|m_*-m'_*|<\epsilon$ we get 
\begin{equation}\label{ineqshellepsilon}
m_*<\frac{2(1+\Omega)m_*\sqrt{1-2m_*/r_*}}{1+\sqrt{1-2m_*/r_*}}+o(\epsilon),
\end{equation}
and solving for $2m_*/r_*$ gives
\begin{equation}\label{ineqpaper1}
\frac{2m_*}{r_*}<\frac{(1+2\Omega)^2-1}{(1+2\Omega)^2}+o(\epsilon).
\end{equation}
Since $\epsilon>0$ is arbitrary this contradicts our assumption on $2m_*/r_*,$ 
which completes the proof of Theorem 1.
\begin{flushright}
$\Box$
\end{flushright}
\textit{Proof of Theorem 2. }
The proof is a direct consequence of the discussion leading to (\ref{eqF}) and 
the formula (\ref{FDs}), cf. also \cite{An1}. Indeed, let 
$$\frac{2M}{R}=\frac{(1+2\Omega)^2-1}{(1+2\Omega)^2}.$$
The formula (\ref{FDs}) with $r_*=R$ and $m_*=M$ gives
\begin{equation}
F_{R}(\nu'_N)=\frac{2M}{(1+2\Omega)\Big(1+\frac{1}{1+2\Omega}\Big)}=\frac{M}{(1+\Omega)},
\end{equation}
and the proof of Theorem 2 is complete.
\begin{flushright}
$\Box$
\end{flushright}
\begin{center}
\textbf{Acknowledgement}
\end{center}
I want to thank Alan Rendall for some comments on the manuscript.

\end{document}